\documentclass[12pt]{article}
\usepackage{subfloat}
\pdfoutput = 1
\usepackage{graphics}
\usepackage{graphicx} 
\usepackage{amsmath}
\usepackage{cite}
\usepackage{amsfonts}   
\usepackage{amssymb}
\usepackage{multirow}
\usepackage{hhline}
\usepackage{color}
\usepackage{comment}
\begin{document}
\date{Today}
\title{{\bf{\Large Non-commutative correction of ideal gas thermodynamics}}}
\author{ {\bf {\normalsize Diganta Parai}
\thanks{digantaparai007@gmail.com}},\,
{\bf {\normalsize Suman Kumar Panja}
\thanks{19phph17@uohyd.ac.in, sumanpanja19@gmail.com}}\\
 {\normalsize School of physics, University of Hyderabad}\\{\normalsize Central University P.O, Hyderabad-500046, Telangana, India}\\[0.2cm] 
\\[0.2cm]
}
\date{}

\maketitle
\begin{abstract}
  In this study, we investigate the thermodynamics of a relativistic ideal within the context of $\kappa$-deformed space-time and Rainbow gravity background. To achieve this, we construct a modified partition function by considering a deformed Hamiltonian and incorporating corrections based on the time-invariant phase-space volume. We explore the implications of our model on the modified black body radiation spectrum and the modified Debye theory of specific heat in $\kappa$-deformed space-time and Rainbow gravity background.
\end{abstract}
\vskip 1cm
\section{Introduction}

At the quantum regime, the nature of gravity is yet unknown and is one of the intriguing unsolved questions in physics. Several theories \cite{connes,douglas,douglas1,rov,sorkin,Glikman,dop,dop1,am,madore,seiberg-witten} like loop quantum gravity, string theory and non-commutative geometry have been proposed to resolve this question. From all these theories it has been observed that there exists an observer independent fundamental length scale associated with each of the theory. This length scale can be identified with Planck length\cite{Maggiore,park}. However this existence of an observer independent minimal length scale is not compatible with special theory of relativity (STR). To resolve this Doubly special relativity(DSR) has been proposed\cite{Camelia,camelia}. In this theory, two invariants are speed of light $c$ and Planck length $l_p$(or Planck energy $E_p$). 

It has been shown that the space-time associated with DSR is $\kappa$-deformed space-time, a non-commutative space-time. In the last couple of decades extensive studies have been made on various non-commutative geometry \cite{wess1,wess2,chaichian,chaichian1,kappa1,dimitrijevic,dasz,mel1,mel2,mel3,carlson,amorim}. $\kappa$-deformed space-time is one where the time and the space coordinates satisfy Lie algebraic type relations, i.e., 
\begin{equation}
\left[X^i,X^j\right]=0~~~\left[X^0,X^i\right]=iaX^i,
\end{equation}   
 where $a$ is the deformation parameter having dimension of length and of the order of Planck length($10^{-35}m$).
 
The generalization of DSR incorporating space-time curvature in the theory is called Rainbow gravity (RG) \cite{ms}. The deformed energy-momentum dispersion relation in RG is given by\cite{ms,Magueijo_Smolin2}
\begin{equation}
E^2f^2(E/E_p)-p^2g^2(E/E_p)=m^2,
\end{equation}
where $m$ and $p$ are the mass and momentum of the test particle. Here $E$ is the energy of the test particle and $E_{p}$ is the Planck energy. The functions $f$ and $g$ are known as rainbow functions. By appropriate choice of these functions one obtains a relation between RG and loop quantum gravity \cite{Alfaro,Sahlmann,smolin}. 

 Different aspects of black hole thermodynamics in $\kappa$-deformed space-time have been studied in \cite{kappa-btz,kappa-btz1,gupta5}. In \cite{kappa-btz,kappa-btz1} corrections to the entropy of BTZ black hole are obtained in the $\kappa$-deformed space-time. Modifications due to the non-commutativity of space-time to Bekenstein-Hawking entropy are obtained in \cite{gupta5}. Study of modifications to Hawking radiation due to $\kappa$-deformed space-time using Bogoliubov coefficients has been addressed in \cite{zuhair1}. To investigate the thermodynamic properties of a system in Rainbow gravity various studies have been conducted \cite{faizal,ali}. In \cite{faizal}, charged dilatonic blackholes thermodynamics is studied in gravity's rainbow. In \cite{ali}, thermodynamic quantities in rainbow Schwarzschild black hole with a particular choice of rainbow functions are calculated. In \cite{cunha}, authors studied the Hawking radiation in rainbow gravity modified Schwarzschild background. Maximum bound on the proper acceleration of a particle is obtained in the Rainbow gravity background in \cite{harsha}. In \cite{ritu}, by incorporating the effects
of the generalized uncertainty principle, the phase transition of a higher dimensional Schwarzschild black hole in the presence of RG is studied.

Modification of thermodynamics in the $\kappa$-deformed space-time and Rainbow gravity background is very interesting to study because it is expected to have quantum gravity effects for ordinary statistical ensemble thermodynamics. Since most physical systems can be idealized by perfect gas fluid or harmonic oscillator ensemble, we will discuss these two systems in the presence of $\kappa$-deformed space-time and Rainbow gravity background in this paper in great detail. This study can be accomplished by two different approaches, namely the modified Hamiltonian method and the modified density of states method\cite{Mirtorabi}. In our paper we consider both approaches to derive the thermodynamical quantities in $\kappa$-deformed space-time and Rainbow gravity. 
 
In the modified Hamiltonian approach, the Hamiltonian of the system is changed by replacing the momentum with generalized momentum and space-time coordinate with corresponding generalized non-commutative coordinates. As a result the traditional Hamiltonian gets additional correction terms. Since the partition function of the system is closely related to the Hamiltonian, the partition function also gets modified. Since the non-commutative parameter is considered to be small, the modified partition function has been evaluated by using the perturbation technique. As an implication of the modified Hamiltonian approach we analyze the Debye specific heat theory in non-commutative set-up. In the second approach, the modified density of states method is used to capture quantum gravity effects from the canonical transformation of the coordinates of phase space. According to Liouville theorem the density of states in a neighbourhood of some phase space point remains constant in time. We will find modified time-invariant phase space volume in the non-commutative set-up. We derive the modified thermodynamical quantities by including modified time-invariant phase space volume in the partition function. 

In this work, we employ both approaches discussed above to derive the modified thermodynamical partition function and compute the thermodynamical quantities for an ensemble of N non-interacting particles moving with relativistic velocities in $\kappa$-deformed space-time and Rainbow gravity background. To observe the thermodynamical implications, we evaluate correction due to a modified Hamiltonian and modification of the time-invariant phase-space volume while calculating the partition function. Using this consideration, we calculate thermodynamical variables such as Helmholtz free energy (F) and Entropy (S) up to the first order in the deformation parameter. We derive the total internal energy (U) for the system in the non-relativistic limit and ultra-relativistic limit. We also study the Black body radiation and find the modification to the Stefan-Boltzmann parameter in $\kappa$-deformed space-time. Further we study Debye theory of specific heat valid up to first order in deformation parameter $a$. We repeat all these studies in the Rainbow gravity background. 

The organization of the paper is as follows. In section-2, we briefly discuss specific realisation in $\kappa$-deformed space-time. Section-3 demonstrates the derivation of time-invariant phase space volume in $\kappa$-deformed background. In section-4, various modified thermodynamical quantities have been derived for relativistic ideal gas in $\kappa$-deformed space-time. In section-5 and -6, we formulated modification of black body radiation spectrum and modification of Debye specific heat theory respectively. In section-7, we give a brief summary of Rainbow gravity. In section-8, we have derived time invariant phase space volume in Rainbow gravity background. In section-9, thermodynamic modifications in rainbow gravity background have been studied. In section-10 and 11 modification of black body radiation spectrum and modified Debye specific heat theory in the Rainbow gravity background have been analysed. In section-12 we give our concluding remarks.

\section{$\kappa$-space-time}

This section summarizes the essential details of the $\kappa$-deformed space-time that is needed for our purposes. $\kappa$-deformed space-time is an example of Lie-algebra type non-commutative space-time and coordinates satisfy Poisson bracket relations
\begin{eqnarray}
\left\{X^i,X^j\right\}&=&0\\
\left\{X^0,X^j\right\}&=&a X^j,
\label{kcommutator0}
\end{eqnarray} 
where $a$ is the deformation parameter with a dimension of length. In the limit $a\rightarrow 0$ we recover the usual commutative space-time. The star product formalism was employed to construct theoretical models of field theory on $\kappa$ space-time. In this formalism, the traditional concept of pointwise product between coordinates (and their functions) is replaced by the star product, which remains unchanged under the $\kappa$-Poincare algebra \cite{dimitrijevic,dasz}. Alternatively, the non-commutative coordinates and their functions can be represented as functions of commutative coordinates and their derivatives using realizations \cite{mel1,mel2}. It has been demonstrated that the realization approach is equivalent to the star product formalism in $\kappa$ space-time \cite{mel3}. For the purpose of this study, we have used the realization approach. One specific realization for non-commutative coordinates and momenta in terms of commutative variables has been discussed in \cite{hari1}. 
In this realization one starts with \cite{hari1} 
\begin{equation}\label{ksp-7}
 X_{\mu}=x_{\nu}\varphi^{\nu}_{\mu},
\end{equation}
where 
\begin{equation}
 \varphi^{\nu}_{\mu}=\delta^{\nu}_{\mu}\Big(1+\alpha(a\cdot p)\Big)+\beta a^{\nu}p_{\mu}+\gamma p^{\nu}a_{\mu}, \label{ksp-7a}
\end{equation}
and $\alpha$, $\beta$, $\gamma$ are constants. Generalizing the commutation relation between the phase-space coordinates to corresponding non-commutative variables (i.e.,$[X_{\mu},P_{\nu}]=i\hbar \hat{g}_{\mu \nu}$) one obtain the canonical momentum operator as \cite{hari1} $P_{\mu}= g_{\alpha \beta}(\hat{y})p^{\alpha}\varphi^{\beta}_{\mu}(p)$. 
This realization takes the form
\begin{eqnarray}
X^{\mu}&=&x^{\mu}+\alpha(a.p)x^{\mu}+\beta(a.x)p^{\mu}+\gamma(x.p) a^{\mu} \nonumber\\
P^{\mu}&=&p^{\mu}+(\alpha+\beta)(a.p)p^{\mu}+\gamma(p.p)a^{\mu},\label{NC1}
\end{eqnarray}
where $\mu$ runs $0,1,2,3$. The fundamental structure of $\kappa$-deformed space-time can be described by the following $X^{\mu}$, $P^{\mu}$ Poisson bracket relations
\begin{eqnarray}
\left\{X^{\mu},X^{\nu}\right\}&=&a^{\mu}X^{\nu}-a^{\nu}X^{\mu}\\
\left\{P^{\mu},P^{\nu}\right\}&=&0\\
\left\{P^{\mu},X^{\nu}\right\}&=&\eta^{\mu\nu}\left[1+s(a.p)\right]+(s+2)a^{\mu}p^{\nu}+(s+1)a^{\nu}p^{\mu},
\end{eqnarray}  
where $s=2\alpha+\beta$. $a^{\mu}=(a,0)$ is the deformation parameter. $\alpha,\beta,\gamma$ satisfy $\gamma-\alpha=1~~~ \beta\in \mathbb{R}$.

\section{ Derivation of invariant phase space volume under time translation}

In this section, we derive the $\kappa$-deformed phase space volume, which is invariant under time transaltion. We use $a^{\mu}=(a,\vec{0})$ in eq.(\ref{NC1}) and find $X^i$ and $P^i$ to be
\begin{equation}
X^i=x^i+\alpha a p^0 x^i+\beta ctap^i
\label{x_rel}
\end{equation}
\begin{equation}
P^i=p^i+(\alpha+\beta)ap^0p^i
\label{p_rel}
\end{equation}
and $X^i$ and $P^i$ satisfy
\begin{equation}
\left\{X^{i},X^{j}\right\}=0=\left\{P^{i},P^{j}\right\}~;~~\left\{P^{i},X^{j}\right\}=-\delta^{ij}\left\{1+(2\alpha+\beta)ap^0\right\},
\label{pxcommutator}
\end{equation}
where $p^0=\sqrt{\vec{p}^2+m^2c^2}$. The time evolution of coordinates and momenta are governed by
\begin{equation}
\dot{X}^i=\left\{X^i,P^j\right\}\frac{\partial H}{\partial P^j}+\frac{\partial X^i}{\partial t}~,~~\dot{P}^i=-\left\{X^j,P^i\right\}\frac{\partial H}{\partial X^j}.
\end{equation}
Now we will find the proper weighted phase space volume invariant under time translation. For calculating it we consider the infinitesimal time interval $\delta t$. In $\delta t$ time coordinates and momenta will be
\begin{equation}
\hat{X}^i=X^i+\delta X^i~;~~\hat{P}^i=P^i+\delta P^i,
\end{equation} 
where
\begin{equation}
\delta X^i=\left[\left\{X^i,P^j\right\}\frac{\partial H}{\partial P^j}+\frac{\partial X^i}{\partial t}\right]\delta t ~,~~\delta P^i=-\left\{X^j,P^i\right\}\frac{\partial H}{\partial X^j}\delta t.
\end{equation}
After this infinitesimal time evolution an infinitesimal phase space volumes are related as
\begin{equation}
d^3\hat{X}d^3\hat{P}=\left|\frac{\partial\left(\hat{X}^1,\hat{X}^2,\hat{X}^3,\hat{P}^1,\hat{P}^2,\hat{P}^3\right)}{\partial\left(X^1,X^2,X^3,P^1,P^2,P^3\right)}\right|d^3Xd^3P.
\label{phasespace}
\end{equation}
Jacobian in the first order of $\delta t$ is given by
\begin{equation}
\left|\frac{\partial\left(\hat{X}^1,\hat{X}^2,\hat{X}^3,\hat{P}^1,\hat{P}^2,\hat{P}^3\right)}{\partial\left(X^1,X^2,X^3,P^1,P^2,P^3\right)}\right|=1+\sum_{i=1}^{3}\left\{\frac{\partial \delta X^i}{\partial X^i}+\frac{\partial \delta P^i}{\partial P^i}\right\}+...
\end{equation}
Thus we find
\begin{equation}
\frac{1}{\delta t}\left[\sum_{i=1}^{3}\left\{\frac{\partial \delta X^i}{\partial X^i}+\frac{\partial \delta P^i}{\partial P^i}\right\}\right]=-\left[\frac{\partial}{\partial P^i}\left\{X^j,P^i\right\}\right]\frac{\partial H}{\partial X^j}.
\label{jacobian}
\end{equation}
Substituting the eq.(\ref{pxcommutator}) in eq.(\ref{jacobian}) we obtain up to first order in deformation parameter $a$ is
\begin{equation}
\frac{1}{\delta t}\left[\sum_{i=1}^{3}\left\{\frac{\partial \delta X^i}{\partial X^i}+\frac{\partial \delta P^i}{\partial P^i}\right\}\right]=-(2\alpha+\beta)a\frac{p^j}{p^0}\frac{\partial H}{\partial X^j}.
\end{equation}
Now eq.(\ref{phasespace}) reads
\begin{equation}
d^3\hat{X}d^3\hat{P}=\left[1-(2\alpha+\beta)\frac{a}{p^0}\left(p^j\frac{\partial H}{\partial X^j}\right)\delta t\right]d^3Xd^3P.
\label{invariant1}
\end{equation}
We will now compute the following quantity
\begin{eqnarray}
1+(2\alpha+\beta)a\sqrt{\hat{P}^2+m^2c^2}&=&1+(2\alpha+\beta)a\sqrt{(P^i+\delta P^i)(P^j+\delta P^j )\delta_{ij}+m^2c^2}\nonumber\\
&=&1+(2\alpha+\beta)a\sqrt{P^2+m^2c^2-2P^i\left\{X^j,P^i\right\}\frac{\partial H}{\partial X^j}\delta t}\nonumber\\
&=&1+(2\alpha+\beta)a\sqrt{P^2+m^2c^2}\left[1-\frac{P^i}{P^2+m^2c^2}\left\{X^j,P^i\right\}\frac{\partial H}{\partial X^j}\delta t\right]\nonumber\\
&=&1+(2\alpha+\beta)a\sqrt{P^2+m^2c^2}\left[1-\frac{P^i}{P^2+m^2c^2}\frac{\partial H}{\partial X^j}\delta t\right]+\mathcal{O}(a^2)\nonumber\\
&=&\left[1+(2\alpha+\beta)a\sqrt{P^2+m^2c^2}\right]\left[1-(2\alpha+\beta)a\frac{P^j}{\sqrt{P^2+m^2c^2}}\frac{\partial H}{\partial X^j}\delta t+\mathcal{O}(a^2)\right]\nonumber\\
&=&\left[1+(2\alpha+\beta)a\sqrt{P^2+m^2c^2}\right]\left[1-(2\alpha+\beta)\frac{a}{p^0}\left(p^j\frac{\partial H}{\partial X^j}\right)\delta t\right]
\label{invariant2}
\end{eqnarray}
Comparing the eq.(\ref{invariant1}) and the eq.(\ref{invariant2}) the time invariant weighted phase space volume becomes
\begin{equation}
\frac{d^3\hat{X}d^3\hat{P}}{1+(2\alpha+\beta)\sqrt{\hat{P}^2+m^2c^2}}=\frac{d^3Xd^3P}{1+(2\alpha+\beta)a\sqrt{P^2+m^2c^2}}
\label{density_states}
\end{equation}   

\section{$\kappa$-deformed  modification to thermodynamics for ideal gas}

To study the effect of $\kappa$-deformed background space-time on the statistical mechanics of a relativistic ideal gas, first, we formulate the modified partition function considering the modified Hamiltonian approach as well as correction due to modified phase space volume has been taken into account. One particle partition function is given by 
\begin{equation}
Q=\frac{1}{h^3}\int\frac{d^3xd^3p}{1+(2\alpha+\beta)a\sqrt{p^2+m^2c^2}}e^{-\frac{H(p)}{kT}},
\label{partition}
\end{equation}
where $k$ is the Boltzmann constant. Using eq.(\ref{p_rel}) we find the modified Hamiltonian as
\begin{equation}
H(p)=mc^2\left[1+\frac{p^2}{m^2c^2}\left\{1+(\alpha+\beta)ap^0\right\}^2\right]^{\frac{1}{2}}-mc^2
\label{Hamiltonian}
\end{equation}
Substituting eq.(\ref{Hamiltonian}) in eq.(\ref{partition}), the single particle partition function read
\begin{equation}
Q=\frac{4\pi V}{h^3}e^u\int_{0}^{\infty}dp p^2\frac{exp\left[-u\sqrt{\left[1+\frac{p^2}{m^2c^2}\left\{1+(\alpha+\beta)amc\sqrt{1+\frac{p^2}{m^2c^2}}\right\}^2\right]}\right]}{1+(2\alpha+\beta)amc\sqrt{1+\frac{p^2}{m^2c^2}}},
\end{equation}
where $u=\frac{mc^2}{kT}$. Now define another variable $s=\frac{p}{mc }$ and also define $A=(\alpha+\beta)amc$ and $B=(2\alpha+\beta)amc$. Reparametrized partition function reads
\begin{equation}
Q=\frac{4\pi V}{h^3}e^u(mc)^3\int_{0}^{\infty} ds~ s^2\frac{e^{-u\sqrt{1+s^2\left\{1+A\sqrt{1+s^2}\right\}^2}}}{1+B\sqrt{1+s^2}}.
\label{Reparametrized}
\end{equation}  
Since $a$ is very small, we can expand the integrand with respect to $a$ and consider up to first order in a
\begin{eqnarray}
\frac{e^{-u\sqrt{1+s^2\left\{1+A\sqrt{1+s^2}\right\}^2}}}{1+B\sqrt{1+s^2}}&\approx & e^{-u\left[1+s^2+2As^2\sqrt{1+s^2}+A^2s^2(1+s^2)\right]^{\frac{1}{2}}}\left(1-B\sqrt{1+s^2}\right)\nonumber\\
&=& e^{-u\sqrt{1+s^2}\left[1+2A\frac{s^2}{\sqrt{1+s^2}}+A^2s^2\right]^{\frac{1}{2}}}\left(1-B\sqrt{1+s^2}\right)\nonumber\\
&\approx & e^{-u\sqrt{1+s^2}\left[1+\frac{As^2}{\sqrt{1+s^2}}\right]}\left(1-B\sqrt{1+s^2}\right)\nonumber\\
&\approx &e^{-u\sqrt{1+s^2}}\left(1-Aus^2\right)\left(1-B\sqrt{1+s^2}\right)\nonumber\\
&\approx & e^{-u\sqrt{1+s^2}}\left\{1-Aus^2-B\sqrt{1+s^2}\right\}.
\end{eqnarray}
Now partition function given in eq.(\ref{Reparametrized}) becomes
\begin{equation}
Q=4\pi V\left(\frac{mc}{h}\right)^3\frac{e^u}{u}K_2(u)\left[1+\frac{B}{u}-(3A+B)\frac{K_3(u)}{K_2(u)}\right],
\end{equation}
where $K_{n}(u)$ is the modified Bessel function of the second kind of integer order $n$. Now, the partition function for N particle is given by
\begin{equation}
Q_N=\frac{Q^N}{N!}=\frac{1}{N!}\left\{4\pi V\left(\frac{mc}{h}\right)^3\frac{e^u}{u}K_2(u)\left[1+\frac{B}{u}-(3A+B)\frac{K_3(u)}{K_2(u)}\right]\right\}^N.
\end{equation}
Using above equation we find Helmholtz free energy as
\begin{equation}
F=-kTlnQ_N=-NkT\left\{ln\left[\frac{4\pi V}{N}\left(\frac{mc}{h}\right)^3\frac{K_2(u)}{u}\right]+u+1\right\}+\Delta F,
\end{equation}
where 
\begin{equation}
\Delta F=NkT\left\{(3A+B)\frac{K_3(u)}{K_2(u)}-\frac{B}{u}\right\}
\end{equation}
is the correction term due to non-commutative space-time. As the pressure $P=-\frac{\partial F}{\partial V}=\frac{NkT}{V}$, the equation of state does not change due to $\kappa$-deformed space-time background. Entropy of the system is given by
\begin{equation}
S=-\frac{\partial F}{\partial T}=Nk\left\{ln\left[\frac{4\pi V}{N}\left(\frac{mc}{h}\right)^3\frac{K_2(u)}{u}\right]+u\frac{K_1(u)}{K_2(u)}+4\right\} +\Delta S,
\end{equation}
where
\begin{equation}
\Delta S=-Nku\left\{(3A+B)\left[1+\frac{2K_3(u)}{uK_2(u)}-\frac{K_1(u)K_3(u)}{K_2^2(u)}\right]-\frac{2B}{u^2 }\right\}
\end{equation}
is the correction term. The total internal energy is given by
\begin{eqnarray}
U=F+TS=NkT\left[3-u\left\{1-\frac{K_1(u)}{K_2(u)}\right\}\right]+\Delta U ,
\label{energy}
\end{eqnarray}
where
\begin{eqnarray}
\Delta U=-NkTu\left\{(3A+B)\left[1+\frac{K_3(u)}{uK_2(u)}-\frac{K_1(u)K_3(u)}{K_2^2(u)}\right]-\frac{B}{u^2 }\right\}
\end{eqnarray}
captures the modification. In the non-relativistic limit when $mc^2>>kT$, i.e, $u>>1$ asymptotic form of modified Bessel function is given by \cite{Mirtorabi}
\begin{equation}
K_n(u)\approx \sqrt{\frac{\pi}{2 u}}e^{-u}\left[1+\frac{4n^2-1}{8u}+\frac{(4n^2-1)(4n^2-9)}{2!(8u)^2}+...\right].
\label{u>>1}
\end{equation} 
Using eq.(\ref{u>>1}) we find
\begin{equation}
\frac{K_1(u)}{K_2(u)}\approx 1-\frac{3}{2u}+\frac{15}{8u^2}+... ~;~~~\frac{K_3(u)}{K_2(u)}\approx 1+\frac{5}{2u}+\frac{15}{8u^2}+...
\label{approx}
\end{equation}
Inserting eq.(\ref{approx}) in eq.(\ref{energy}) we find the total internal energy as
\begin{eqnarray}
U&=&\frac{3}{2}NkT\left[1-\frac{5A+B}{u}\right]\\
&=&\frac{3}{2}NkT\left[1-(7\alpha+6\beta)a\left(\frac{kT}{c}\right)\right].
\end{eqnarray}
For ultrarelativistic limit when $mc^2<<kT$ i.e, $u\rightarrow 0$ form of modified Bessel function is given by\cite{Mirtorabi}
\begin{equation}
K_n(u)\approx \frac{1}{2}(n-1)!\left(\frac{u}{2}\right)^{-n}.
\label{u<<1}
\end{equation} 
Inserting eq.(\ref{u<<1}) in eq.(\ref{energy}) we find total internal energy to be
\begin{eqnarray}
U&=&3NkT\left[1-\frac{4A+B}{u}\right]\\
&=&3NkT\left[1-(6\alpha+5\beta)a\left(\frac{kT}{c}\right)\right]
\end{eqnarray}

\section{Modified black body radiation spectrum}
It is interesting to study the black body radiation spectrum modification due to the $\kappa$-deformed background space-time. According to the eq.(\ref{density_states}) the modified density of states will be
\begin{equation}
g(p)dp=\frac{8 \pi V}{h^3}\frac{p^2dp}{1+(2\alpha+\beta)ap},
\end{equation}
where we have considered the rest mass energy of the photon to be zero. Using the relation $p=\frac{h\nu}{c}$ we find
\begin{equation}
g(\nu)d\nu=\frac{8\pi V}{c^3}\frac{\nu^2d\nu}{1+(2\alpha+\beta)a\frac{h\nu}{c}}.
\end{equation} 
The statistical distribution function for photon is given by
\begin{equation}
N(\nu)d\nu=\frac{g(\nu)d\nu}{e^{\frac{h\nu}{kT}}-1}.
\label{statistics}
\end{equation}
Average energy of electromagnetic field per unit volume at temperature $T$ is given by
\begin{eqnarray}
U(T)&=&\int_{0}^{\infty}h\nu N(\nu)d\nu\nonumber\\
&=&\frac{8\pi h}{c^3}\int_{0}^{\infty}\frac{\nu^3d\nu}{\left(1+\frac{\nu}{\nu_0}\right)\left(e^{\frac{h\nu}{kT}}-1\right)},
\label{avarage_energy}
\end{eqnarray}
where $\nu_0^{-1}=(2\alpha+\beta)\frac{ah}{c}$. Expanding the integrand of eq.(\ref{avarage_energy}) in the first order of deformation parameter $a$ we find
\begin{equation}
U(T)= \frac{8\pi h}{c^3}\int_{0}^{\infty}\left(1-\frac{\nu}{\nu_0}\right)\frac{\nu^3d\nu}{\left(e^{\frac{h\nu}{kT}}-1\right)}.
\label{approx_avg_energy}
\end{equation}
Using the identity involving Riemann Zeta function $\xi(\nu)$
\begin{equation}
\xi(\nu)=\frac{1}{\Gamma(\nu)}\int_{0}^{\infty}dx\frac{x^{\nu-1}}{e^x-1}
\label{Riemann_zeta}
\end{equation}
we evaluate eq.(\ref{approx_avg_energy}) and after calculational simplifications we finally find
\begin{equation}
U(T)=\frac{8 \pi^5k^4T^4}{15h^3c^3}\left[1-\frac{360}{\pi^2}\xi(5)\frac{T}{T_{\nu}}\right],
\end{equation}
where $T_{\nu}=\frac{h\nu_0}{k}=\frac{c}{(2\alpha+\beta)ka}$. Hence radiation spectra becomes
\begin{equation}
U(T)=\frac{4}{c}\sigma_{eff}T^4,
\end{equation} 
where
\begin{equation}
\sigma_{eff}=\frac{2\pi^5k^4}{15c^2h^3}\left[1-\frac{360}{\pi^2}\xi(5)\frac{T}{T_{\nu}}\right]
\label{sigma}
\end{equation}
is the effective Stefan-Boltzmann parameter. The first term of eq.(\ref{sigma}) gives the ordinary Stefan-Boltzmann constant whereas the second temperature-dependent term signifies $\kappa$-deformed correction term. Note that here we have taken modifications valid up to first order in deformation parameter $a$. The above expression of radiation spectra reduce to the commutative result in the limit $a \rightarrow 0$.

\section{$\kappa$-deformed Debye specific heat theory}
As an application of statistical outcome considered in $\kappa$-deformed space-time, it is interesting to study the modification of Debye specific heat theory. Like in Debye theory we will consider a solid N-particle system represented as a three dimensional array of coupled spring-mass systems. First we will describe a single spring mass system in $\kappa$-deformed space-time. The modified Hamiltonian is given by
\begin{equation}
H=\frac{P^2}{2m}+\frac{1}{2}m \omega^2 X^2,
\end{equation}
where noncommutative coordinate $X$ and momenta $P$ are related to ordinary coordinate and momenta as
\begin{equation}
X^i=(1+\alpha a p^0)x^i~\text{and}~~P^i=\left(1+\alpha ap^0\right)p^i,
\end{equation}
respectively. Here we have considered $\beta=0$ in eq.(\ref{x_rel}) and eq.({\ref{p_rel}}). Now modified Hamiltonian in terms of commutative coordinates and momenta up to first order in deformation parameter read
\begin{equation}
H=\left(1+2\alpha a p^0\right)\left[\frac{p^2}{2m}+\frac{1}{2}m\omega^2 x^2\right].
\end{equation}
So the expression for allowed quantum energy levels are
\begin{equation}
E_n=\left(n+\frac{1}{2}\right)h\nu(1+2\alpha a p^0)\equiv\left(n+\frac{1}{2}\right)h\nu q, ~~~q\equiv (1+2\alpha a p^0).
\end{equation}
Now we use the classical result for a three dimensional cubic array of N-particle coupled spring-mass system. The number of normal modes  in the range $\nu$ to $\nu+d\nu$ is given by
\begin{equation}
g(\nu)d\nu=\frac{4\pi V}{\xi^3}\nu^2d\nu,
\end{equation}
where $\xi=\frac{v_s}{3^{\frac{1}{3}}}$, $v_s$ is the sound velocity in the solid and $V$ is the volume of solid. Since $\int_{0}^{\nu_D}g(\nu)d\nu=3N$, where $\nu_D$ is the Debye frequency, sets the condition
\begin{equation}
\frac{9N}{\nu_D^3}=\frac{4\pi V}{\xi^3}
\end{equation}
Average energy at temperature $T$ of the system is given by
\begin{eqnarray}
U(T)&=&\int_{0}^{\nu_D}d\nu\frac{h q g(\nu)}{e^{\frac{h q \nu}{kT}}-1}\nonumber\\
&=&\frac{9 N h q}{\nu_D^3}\int_{0}^{\nu_D}d\nu\frac{\nu^3}{e^{\frac{h q \nu}{kT}}-1}.
\end{eqnarray}
Changing the variable $\tilde{\nu}=q\nu$ we find the above integration as
\begin{eqnarray}
U(T)=\frac{9Nh}{\tilde{\nu}_D^3}\int_{0}^{\tilde{\nu}_D}d\tilde{\nu}\frac{\tilde{\nu}^3}{e^{\frac{h\tilde{\nu}}{kT}}-1}.
\label{dedye_energy}
\end{eqnarray}
For $T<<\tilde{\theta}_D=\frac{h\tilde{\nu}_D}{k}=(1+2\alpha a p^0)\frac{h\nu_D}{k}$ we find specific heat $C_V$ from eq.(\ref{dedye_energy}) as
\begin{equation}
C_V(T)=\frac{12}{5}\pi^4 Nk\left(\frac{T}{\tilde{\theta}_D}\right)^3,
\end{equation}
which is well known Debye $T^3$ law. From this analysis we observe that the Debye temperature is modified given by
\begin{equation}
\tilde{\theta}_D=(1+2\alpha a p^0)\frac{h\nu_D}{k}.
\end{equation}
In all the calculations, we have taken correction up to the first order in $a$. In $a\rightarrow 0$ we recover the usual commutative result. 

\section{Rainbow gravity}
Various class of theories has been developed in the background of Doubly special relativity (DSR)\cite{Camelia,camelia}. DSR is based upon three postulates namely, (i) special relativity postulates should hold in all inertial frames, (ii) in the limit $E/E_{pl}\rightarrow 0$ speed of photon is a universal constant $c$ which is the same for all inertial frames and (iii) in all inertial frame $E_{pl}$ is also a universal constant. Rainbow gravity\cite{ms,Magueijo_Smolin2} is one of the developed theory in DSR. Energy momentum dispersion relation in rainbow gravity is given by
\begin{equation}
E^2f^2(E)-\textbf{p}^2c^2g^2(E)=m^2c^4.
\end{equation}
This particular realization comes from the action of a non-linear map from momentum space to itself given by\cite{ms}
\begin{equation}
U.(E/c,p^i)=(f(E)E/c,g(E)p^i).
\end{equation} 
Here we will consider $f(E)=g(E)=\frac{1}{1-\lambda p^0 }$\cite{ms}. Where $\lambda$ is the rainbow parameter which has dimension of length. Now dispersion relation reads
\begin{equation}
\frac{\eta_{\mu\nu}p^{\mu}p^{\nu}}{(1-\lambda p^0)^2}=m^2c^2.
\end{equation}
Using the same choice of rainbow function coordinate variable transforms as\cite{Kimberly}
\begin{equation}
X^i=(1-\lambda p^0)x^i.
\label{x_relation}
\end{equation}

\section{Derivation of time-invariant phase space volume for Rainbow gravity}
In this section, we derive invariant phase space volume for rainbow gravity. We start with modified position $X^i$ and $P^j$ satisfy the Poisson bracket relations\cite{Magueijo_Smolin2}
\begin{equation}
\left\{X^i,X^j\right\}=0=\left\{P^i,P^j\right\}
\end{equation}
\begin{equation}
\left\{X^i,P^j\right\}=\delta^{ij}(1-\lambda P^0),
\label{rainbow_rel}
\end{equation}
where $P^0=\sqrt{\vec{P}^2+m^2c^2}$. The time evolution of modified position and momentum coordinates are governed by
\begin{equation}
\dot{X}^i=\left\{X^i,P^j\right\}\frac{\partial H}{\partial P^j}~\text{and}~~\dot{P}^i=-\left\{X^j,P^i\right\}\frac{\partial H}{\partial X^j}.
\end{equation}
To calculate proper phase space volume invariant under time translation, first we write infinitesimal time translated coordinate as
\begin{equation}
\hat{X}^i=X^i+\delta X^i ~;~~\hat{P}^i=P^i+\delta P^i ,
\end{equation}
where
\begin{equation}
\delta X^i=\left\{X^i,P^j\right\}\frac{\partial H}{\partial P^j}\delta t ~;~~ \delta P^i=-\left\{X^j,P^i\right\}\frac{\partial H}{\partial X^j}\delta t.
\end{equation}
After this infinitesimal time evolution, infinitesimal phase space volume is given by
\begin{equation}
d^3\hat{X}d^3\hat{P}=\left|\frac{\partial\left(\hat{X}^1,\hat{X}^2,\hat{X}^3,\hat{P}^1,\hat{P}^2,\hat{P}^3\right)}{\partial\left(X^1,X^2,X^3,P^1,P^2,P^3\right)}\right|d^3Xd^3P.
\label{phasespace2}
\end{equation}
Jacobian of this transformation in the first order of $\delta t$ is given by
\begin{equation}
\left|\frac{\partial\left(\hat{X}^1,\hat{X}^2,\hat{X}^3,\hat{P}^1,\hat{P}^2,\hat{P}^3\right)}{\partial\left(X^1,X^2,X^3,P^1,P^2,P^3\right)}\right|=1+\sum_{i=1}^{3}\left\{\frac{\partial \delta X^i}{\partial X^i}+\frac{\partial \delta P^i}{\partial P^i}\right\}+...
\end{equation}
 Using eq.(\ref{rainbow_rel}) we find,
\begin{equation}
\frac{1}{\delta t}\left[\sum_{i=1}^{3}\left\{\frac{\partial \delta X^i}{\partial X^i}+\frac{\partial \delta P^i}{\partial P^i}\right\}\right]=-\left[\frac{\partial}{\partial P^i}\left\{X^j,P^i\right\}\right]\frac{\partial H}{\partial X^j}=\lambda\frac{P^i}{P^0}\frac{\partial H}{\partial X^i}\delta t.
\label{jacobian2}
\end{equation}
Now eq.(\ref{phasespace2}) read
\begin{equation}
d^3\hat{X}d^3\hat{P}=\left[1+\lambda\frac{P^i}{P^0}\frac{\partial H}{\partial X^i}\delta t\right]d^3Xd^3P.
\label{rainbow_phasespace1}
\end{equation}
Now we calculate the following quantity
\begin{eqnarray}
1-\lambda\hat{P}^0&=&1-\lambda\sqrt{(P^i+\delta P^i)(P^j+\delta P^j)\delta_{ij}+m^2c^2}\nonumber\\
&= &1-\lambda\sqrt{P^2+m^2c^2-2P^i\left\{X^j,P^i\right\}\frac{\partial H}{\partial X^j}\delta t}\nonumber\\
&=&1-\lambda\sqrt{P^2+m^2c^2}\left[1-\frac{P^i}{(P^0)^2}\left\{X^j,P^i\right\}\frac{\partial H}{\partial X^j}\delta t\right]\nonumber\\
&=&1-\lambda\sqrt{P^2+m^2c^2}+\lambda\frac{P^i}{P^0}\frac{\partial H}{\partial X^i}\delta t+\mathcal{O}(\lambda^2)\nonumber\\
&=&(1-\lambda\sqrt{P^2+m^2c^2})\left\{1+\lambda\frac{P^i}{P^0}\frac{\partial H}{\partial X^i}\delta t+\mathcal{O}(\lambda^2)\right\}\nonumber\\
&=&\left(1-\lambda P^0\right)\left\{1+\lambda\frac{P^i}{P^0}\frac{\partial H}{\partial X^i}\delta t\right\}
\label{rainbow_phasespace2}
\end{eqnarray}
Comparing eq.(\ref{rainbow_phasespace1}) with eq.(\ref{rainbow_phasespace2}) we get the time invariant weighted phase space element as
\begin{equation}
\frac{d^3\hat{X}d^3\hat{P}}{1-\lambda\hat{P}^0}=\frac{d^3Xd^3P}{1-\lambda P^0}.
\label{rainbow_density_states}
\end{equation}

\section{Thermodynamical modification for an ideal gas in rainbow gravity background} 
In this section, we study the thermodynamic quantities in the background of Rainbow gravity. For this we start with the Dispersion relation in Rainbow gravity background \cite{ms,Magueijo_Smolin2}
\begin{equation}
\frac{\eta_{ab}p^ap^b}{(1-\lambda p^0)^2}=m^2c^2.
\label{dispersion}
\end{equation}
Writing $p^0=\frac{E}{c}$, from the above equation we find energy expression as
\begin{equation}
E=\frac{mc^2}{1-m^2c^2\lambda^2}\left\{1+(1-m^2c^2\lambda^2)\frac{p^2}{m^2c^2}\right\}^{\frac{1}{2}}-\frac{\lambda m^2 c^3}{1-m^2c^2\lambda^2}.
\label{modified_energy}
\end{equation}
Thus from eq.(\ref{modified_energy}) we find modified Hamiltonian for free particle as
\begin{equation}
H(p)=\frac{mc^2}{1-m^2c^2\lambda^2}\left\{1+(1-m^2c^2\lambda^2)\frac{p^2}{m^2c^2}\right\}^{\frac{1}{2}}-\frac{\lambda m^2 c^3}{1-m^2c^2\lambda^2}-mc^2.
\end{equation}
Now we will construct the single particle partition function considering the modified Hamiltonian and the modified phase space volume element. Single particle partition function is given by
\begin{eqnarray}
Q&=&\frac{1}{h^3}\int \frac{d^3xd^3p}{1-\lambda p^0}\,e^{-\frac{H(p)}{kT}}\nonumber\\
&=&\frac{4\pi V}{h^3}e^ue^{\frac{uA\xi_1}{1-A^2\xi_1}}\int_{0}^{\infty}\frac{p^2 dp}{1-A\xi_2\sqrt{1+\frac{p^2}{m^2c^2}}} e^{-\frac{u}{1-A^2\xi_1}\left\{1+(1-A^2\xi_1)\frac{p^2}{m^2c^2}\right\}^{\frac{1}{2}}},
\label{rainbow_partition}
\end{eqnarray} 
where $u=\frac{mc^2}{kT}$, $A=\lambda m c$ and $\xi_1$ and $\xi_2$ are bookkeeping terms for keeping track of the contribution coming from modified Hamiltonian or corrected phase space volume. After making a variable change $s=\frac{p}{mc}$ eq.(\ref{rainbow_partition}) read
\begin{equation}
Q=\frac{4\pi V}{h^3}e^ue^{\frac{uA\xi_1}{1-A^2\xi_1}}(mc)^3\int_{0}^{\infty}\frac{s^2ds}{1-A\xi_2\sqrt{1+s^2}}e^{-\frac{u}{1-A^2\xi_1}\left\{1+(1-A^2\xi_1)s^2\right\}^{\frac{1}{2}}}.
\end{equation}
As $\lambda$ is of the Planck length order, $A=\lambda mc$ is very small we can neglect $A^2$ term and we get
\begin{eqnarray}
Q&=&\frac{4\pi V}{h^3}(1+uA\xi_1)\,e^u\,(mc)^3\int_{0}^{\infty}ds~s^2\left(1+A\xi_2\sqrt{1+s^2}\right)e^{-u\sqrt{1+s^2}}\nonumber\\
&=&\frac{4\pi V}{h^3}(1+uA\xi_1)\,e^u\,(mc)^3\left[\frac{K_2(u)}{u}+A\xi_2\frac{K_3(u)}{u}-A\xi_2\frac{K_2(u)}{u^2}\right].
\end{eqnarray} 
Now, the N particle partition function is given by
\begin{equation}
Q_N=\frac{Q^N}{N!}=\frac{1}{N!}\left[4\pi v\left(\frac{mc}{h}\right)^3\frac{e^u}{u}K_2(u)\left\{(1+uA\xi_1)+A\xi_2\frac{K_3(u)}{K_2(u)}-\frac{A\xi_2}{u}\right\}\right]^N
\end{equation}
Next we derive the Helmholtz free energy. By definition Helmholtz free energy is given by
\begin{equation}
F=-kTlnQ_N=-NkT\left\{ln\left[\frac{4\pi V}{N}\left(\frac{mc}{h}\right)^3\frac{K_2(u)}{u}\right]+u+1\right\}+\Delta F ,
\end{equation} 
where
\begin{equation}
\Delta F=NkT\left\{\frac{A\xi_2}{u}-uA\xi_1-A\xi_2\frac{K_3(u)}{K_2(u)}\right\}
\end{equation}
is the correction term. Pressure is given by $P=\frac{\partial F}{\partial V}=\frac{NkT}{V}$. So the equation of state does not get modified. Entropy of the system is given by
\begin{equation}
S=-\frac{\partial F}{\partial T}=Nk\left\{ln\left[\frac{4\pi V}{N}\left(\frac{mc}{h}\right)^3\frac{K_2(u)}{u}\right]+u\frac{K_1(u)}{K_2(u)}+4\right\} +\Delta S ,
\end{equation}
where
\begin{equation}
\Delta S=-Nku\left\{\frac{2A\xi_2}{u^2}-A\xi_2\left[1+\frac{2K_3(u)}{uK_2(u)}-\frac{K_1(u)K_3(u)}{K_2^2(u)}\right]\right\}
\end{equation}
is the correction term. Note that in the above no corrections are present due to the modification of the Hamiltonian. Total internal energy is given by
\begin{equation}
U=F+TS=NkT\left[3-u\left\{1-\frac{K_1(u)}{K_2(u)}\right\}\right]+\Delta U ,
\end{equation}
where
\begin{equation}
\Delta U=-NkTu\left\{\frac{A\xi_2}{u^2}-A\left((\xi_2-\xi_1)+\xi_2\frac{K_3(u)}{uK_2(u)}-\xi_2\frac{K_1(u)K_3(u)}{K_2^2(u)}\right)\right\}
\end{equation}
is the modification term. In the non-relativistic limit, i.e, $u>>1$ using eq.(\ref{approx}) we get
\begin{equation}
U=\frac{3}{2}NkT\left\{1-\frac{2\xi_1}{3}\frac{\lambda m^2c^3}{kT}+\xi_2\lambda\frac{kT}{c}\right\}.
\end{equation}
Immediately we can calculate the specific heat constant volume is given by
\begin{equation}
C_V=\frac{\partial U}{\partial T}=\frac{3}{2}Nk\left\{1+2\lambda\xi_2\frac{kT}{c}\right\}.
\end{equation}
For ultrarelativistic limit $u<<1$ using eq.(\ref{u<<1}) we get
\begin{equation}
U=3NkT\left\{1+\lambda\xi_2\frac{kT}{c}\right\}
\end{equation}
and in this limit specific heat at constant volume is given by
\begin{equation}
C_V=3Nk\left\{1+2\lambda\xi_2\frac{kT}{c}\right\}.
\end{equation}
Note here all the modifications to thermodynamic variables are taken up to first order in Rainbow parameter $\lambda$. Note that correction terms in both cases increase the value of specific heat for the non-relativistic and ultra-relativistic limits. This contradict the result we obtained in case of $\kappa$-deformed space-time earlier in Section-4.

\section{Modified Black body radiation and Debye specific heat theory}
In this section, we investigate the Black body radiation in the Rainbow gravity background. We also studied the Debye specific heat theory and found the modified specific heat at constant volume. Modified density of states according to eq.(\ref{rainbow_density_states}) is
\begin{equation}
g(p)dp=\frac{8\pi V}{h^3}\frac{p^2dp}{1-\lambda p^0},
\end{equation}
where mass of the photon is assumed to be zero. Using relation $p^0=\frac{h\nu}{c}$ we find
\begin{equation}
g(\nu)d\nu=\frac{8\pi V}{c^3}\frac{\nu^2d\nu}{1-\lambda\frac{h\nu}{c}}=\frac{8\pi V}{c^3}\frac{\nu^2d\nu}{1-\frac{\nu}{\nu_0}},
\end{equation} 
where we define $\nu_0^{-1}=\frac{\lambda h}{c}$. Using eq.(\ref{statistics}) we can write average energy density of the electromagnetic field at temperature $T$ is given by
\begin{equation}
U(T)=\frac{8\pi h}{c^3}\int_{0}^{\infty}\frac{\nu^3d\nu}{\left(1-\frac{\nu}{\nu_0}\right)\left(e^{\frac{h\nu}{kT}}-1\right)}\approx\frac{8\pi h}{c^3}\int_{0}^{\infty}\left(1+\frac{\nu}{\nu_0}\right)\frac{\nu^3d\nu}{\left(e^{\frac{h\nu}{kT}}-1\right)}
\end{equation} 
Using Riemann zeta identity eq.(\ref{Riemann_zeta}) we can integrate above equation and finally get
\begin{equation}
U(T)=\frac{8 \pi^5k^4T^4}{15h^3c^3}\left[1+\frac{360}{\pi^2}\xi(5)\frac{T}{T_{\nu}}\right],
\end{equation}
where $T_{\nu}=\frac{h\nu_0}{k}$. So, the effective Stefan-Boltzmann parameter is given by
\begin{equation}
\sigma_{eff}=\frac{2\pi^5k^4}{15c^2h^3}\left[1+\frac{360}{\pi^2}\xi(5)\frac{T}{T_{\nu}}\right].
\end{equation}
Here we have considered modification valid up to first order in Rainbow parameter $\lambda$. We observe that the Stefan-Boltzmann parameter increases due to the modification coming from Rainbow gravity. In the limit $\lambda \rightarrow 0$, we get back the usual form of the $\sigma_{eff}$.
 
Next we construct and analyze the Debye specific heat theory in Rainbow gravity background. For this we use eq.(\ref{modified_energy}) with eq.(\ref{x_relation}) and we find 
the modified Hamiltonian for harmonic oscillator in terms of commutative coordinates as
\begin{equation}
H=\frac{p^2}{2m}+\frac{1}{2}m\tilde{\omega}^2x^2 ,
\end{equation}
where $\tilde{\omega}=(1-\lambda p^0)$. So allowed quantum energy levels are
\begin{equation}
E_n=\left(n+\frac{1}{2}\right)h\nu(1-\lambda p^0)
\end{equation}
By following the procedure of calculation of the specific heat given in Section-6, here also we find
\begin{equation}
C_V(T)=\frac{12}{5}\pi^4Nk\left(\frac{T}{\tilde{\theta}_D}\right)^3
\end{equation}
for $T<<\tilde{\theta}_D=(1-\lambda p^0)\frac{h\nu_D}{k}$ , where $\tilde{\theta}_D$ is the modified Debye frequency. Note that contribution due to the Rainbow gravity decreases the Debye frequency. In the limit $\lambda \rightarrow 0$, we get back usual result.
\section{Conclusions}
In this paper, we studied the effect of $\kappa$-deformed space-time and Rainbow gravity background on various thermodynamical quantities for a canonical ensemble of N noninteracting particles moving with relativistic velocities. To observe the thermodynamical consequences here we have considered correction due to the modified Hamiltonian and modification of the time-invariant phase-space volume, for calculating the partition function. Using this consideration, we have computed various modified thermodynamic quantities in linear order in the deformation parameter ($a$). We have studied modified thermodynamical quantities for $\kappa$-deformed space-time and for Rainbow gravity background. It has been observed that for $\kappa$-deformed space-time consideration, we get a negative correction term for modified internal energy in both non-relativistic and ultrarelativistic limits, whereas for Rainbow gravity background, it is positive in both the limits. We notice that specific heat in the Rainbow gravity background has no effect from the modified Hamiltonian approach, correction term arises solely from modified phase-space consideration.

As an implication of our model we have studied black body radiation spectra and Debye theory of specific heat in both backgrounds. In $\kappa$-deformed space-time, effective Stefan-Boltzmann constant has been decreased and for the case of Debye theory we have observed  $T^3$ law has been followed with a modified Debye temperature which is increased due to the deformation parameter. The analysis of two classical theories mentioned above in the Rainbow gravity background shows that the effective Stefan-Boltzmann constant has increased and the modified Debye temperature has decreased, followed by the general $T^3$ law.  

\section*{Acknowledgments}
DP and SKP thank E. Harikumar for the useful discussion. DP thanks IOE-UOH for support through the PDRF scheme. SKP thanks UGC, India, for the support through the JRF scheme (id.191620059604).

\end{document}